\documentclass[12pt]{iopart}
\usepackage[]{graphicx}
\begin{document}

\title[Generation of entangled photons with a second-order nonlinear photonic crystal]{Generation of entangled photons with a second-order nonlinear photonic crystal and a beam splitter}

\author{Hiroo Azuma\footnote{Present address: Global Research Center for
Quantum Information Science, National Institute of Informatics, 2-1-2 Hitotsubashi, Chiyoda-ku, Tokyo 101-8430, Japan}}

\address{Quemix Inc., 16F Taiyo Life Nihonbashi Building, 2-11-2 Nihonbashi, Chuo-ku, Tokyo 103-0027, Japan}
\ead{hiroo.azuma@m3.dion.ne.jp}
\vspace{10pt}
\begin{indented}
\item[]May 2022
\end{indented}

\begin{abstract}
We discuss the generation of entangled photons using a nonlinear photonic crystal and beam splitter.
In our method,
the photonic crystal is assumed to be composed of a material with a large second-order nonlinear optical susceptibility $\chi^{(2)}$.
Our proposal relies on two facts:
(1) A nonlinear photonic crystal changes coherent incident light into squeezed light.
(2) A beam splitter transforms the squeezed light into entangled light beams flying in two different directions.
We estimate the yield efficiency of pairs of entangled photons per pulse of the very weak coherent light for our method at $0.0783$ for a specific concrete example.
Our method is more effective because the conversion efficiency
(entangled biphotons per incident pump photons)
in the spontaneous parametric downconversion is of the order of $4\times 10^{-6}$.
The only drawback is that it requires very fine tuning of the frequency of signal photons fed into the photonic crystal;
for example, an adjustment is given by $\Delta\nu=3.11\times 10^{8}$ Hz for the signal light of $\nu=3.23\times 10^{14}$ Hz.
We investigate the application of entangled photons produced by our method to the BB84 quantum key distribution protocol, and we explore how to detect an eavesdropper.
We suggest that our method is promising for decoy-state quantum key distribution
using weak coherent light.
\end{abstract}

%

%
\submitto{\JPD}
%
%
%

\section{\label{section-Introduction}Introduction}
The development of an on-demand single-photon
source is key for realizing quantum information processing \cite{Nielsen2000}.
Researchers have presented several proposals to develop an on-demand single-photon gun; for example,
using cavity quantum electrodynamics \cite{Kuhn1999,Brattke2001,Mucke2013},
quantum dots \cite{Benson2000,Michler2000,Santori2001,Pelton2002,Kiraz2004},
separate nitrogen vacancy centers in diamond \cite{Bernien2012},
a negative silicon vacancy center in diamond \cite{Rogers2014},
and nitrogen vacancy centers in 4H-SiC
\cite{Zargaleh2016,Wang2017}.
Recently, the emission of single photons at room temperature has been studied using cubic silicon carbide \cite{Wang2018},
strongly interacting Rydberg atoms \cite{Ripka2018},
and a color center in two-dimensional hexagonal boron nitride \cite{Dietrich2020}.
The realization of an ideal single-photon source is
challenging, and it is yet to be accomplished.

We explain the difficulty in developing an on-demand single-photon gun based on the following considerations:
We use weak coherent light as a substitute for the single-photon source.
The number of photons and time interval of successive pulses are random because photons constituting coherent light exhibit a Poisson distribution.
That is, the bunching effect induced by Bose--Einstein statistics makes it difficult to manipulate single photons.

An on-demand single-photon source is a key component in quantum computation,
quantum communication, and quantum cryptography.
For example, the
\\
Knill--Lafalamme--Milburn method used for operating a two-qubit gate requires deterministic multiple single-photon guns \cite{Knill2001,Ralph2001,Kok2007}.
Quantum communication processes, which include quantum teleportation, provide procedures for sending and receiving quantum states of single photons \cite{Bennett1993}.
Therefore, 
many processes of quantum communication postulate an on-demand single-photon gun.
The BB84 quantum key distribution protocol requires a single-photon source to avoid interference from eavesdroppers \cite{Bennett1984,Bennett1992}.

We utilize a pair of entangled photons as an alternative plan for emitting single photons at a specific time. A typical method to realize this is spontaneous parametric downconversion (SPDC).
In SPDC,
a photon with wavelength $\lambda$ is placed into a nonlinear crystal and a pair of photons with wavelengths $2\lambda$ is emitted in two different directions.
If we detect one photon as the heralding signal,
the other one is emitted at the time of the detector's measurement.
This phenomenon is considered a practical process for a deterministic single-photon source.

SPDC is an excellent process that can be used as the principle for generating a single-photon on-demand.
However, its disadvantage is that the conversion efficiency of the entangled photons is considerably low.
In \cite{Schneeloch2016}, the conversion efficiency of an SPDC with a beta barium borate (BBO) crystal
(biphotons per incident pump photon)
is of the order $10^{-8}$.
An SPDC experiment with a BBO crystal was reported in reference~\cite{Nambu2002}, and its rate of coincident counts of entangled pairs was $450$ cps.
In the latest experiment,
the conversion efficiency of the entangled pairs was $4\times 10^{-6}$ per pump photon at most when using a periodically poled lithium niobate (PPLN) waveguide for the SPDC \cite{Bock2016}.
The low efficiency of SPDC hinders the realization of high capacity optical transmission.

We discuss the generation of entangled photons with a one-dimensional photonic crystal
composed of a material with a large second-order nonlinear optical susceptibility $\chi^{(2)}$
and beam splitter.
This method exhibits a considerably higher conversion efficiency of entangled photons than SPDC.
Our proposal relies on the following two facts:

1) A one-dimensional photonic crystal with nonlinear susceptibility $\chi^{(2)}$
transforms coherent incident light into squeezed light \cite{Sakoda2005}.
Then, the photonic crystal allows the group velocity of the incident light to decrease.
If the photonic crystal is composed of a nonlinear material with susceptibility $\chi^{(2)}$,
the slow velocity of light strengthens the interaction between photons and the nonlinear material.
Thus, this process converts the coherent light into squeezed light efficiently.

2) The beam splitter splits the flow of the squeezed light into a pair of entangled light beams \cite{Kim2002}.
In particular, the beam splitter transforms a single-mode squeezing operator applied to the incident light
into a product of single-mode and two-mode squeezing operators.
This two-mode squeezing operator makes two outgoing beams from the beam splitter entangled with each other.
After the two entangled beams travel in two different directions from the beam splitter, each beam loses the bunching effect and the probability of a state of a single photon increases.

Although these two facts are known in the field of quantum optics, they are yet to be combined and reported in academic journals or confirmed via experiments.
(Vamivakas {\it et al} considered the SPDC process in a photonic crystal but they did not discuss it
in the context of squeezing \cite{Vamivakas2004}.)

The proposed method provides a very effective setup for generating entangled pairs of photons.
In this paper, we analyze a concrete example of our proposal using a nonlinear material, lithium niobate LiNbO${}_{3}$.
In this example, the rate of the generation of a pair of entangled photons is estimated to be $0.0783$ per pulse,
which is extremely large compared to the conversion efficiency of the SPDC, which is $4\times 10^{-6}$ per pump photon at most.
Hence, our method is superior to the SPDC process in terms of the yield of entangled photons.

More specifically, in the example,
we inject the coherent light $|\alpha\rangle$ for $\alpha=1/2$ into the nonlinear photonic crystal as a signal beam and obtain a pair of entangled photons
with a probability of $0.0783$.
Thus, the conversion efficiency of the entangled pair is $0.313$ per signal photon.
In our method, as explained in section~\ref{section-photonic-crystal-parameters},
if we assume that the radiant flux of the pump light and the repetition rate of the incident signal pulse are given by $0.03$ W and $10$ MHz, respectively,
the efficiency of entangled photons is estimated at $7.83\times 10^{5}$ pairs per second.

In references~\cite{Somaschi2016,Senellart2017},
brightness is proposed as one of the important properties that describe an ideal single-photon source.
The brightness of the source implies the number of photons collected per excitation pulse.
According to this definition, our scheme provides the brightness $0.0783$.
In reference~\cite{Somaschi2016}, it is shown that the typical brightness of SPDC is lower than $0.08$.
However, to obtain high brightness for SPDC,
we need to increase the power of the pump laser light;
for example, $0.15$ W was required for only $450$ entangled pairs per second in reference~\cite{Nambu2002}.
In the experiments of \cite{Nambu2002}, the repetition rate was $82$ MHz, so that the brightness was given by $5.5\times 10^{-6}$.

The SPDC can generate entangled photons in different modes depending on phase matching.
The generation of squeezed light with the photonic crystal requires a phase-matching condition,
$\mbox{\boldmath $k$}_{\mbox{\scriptsize p}}
=
2\mbox{\boldmath $k$}_{\mbox{\scriptsize s}}
+
\mbox{\boldmath $G$}$,
where $\mbox{\boldmath $k$}_{\mbox{\scriptsize p}}$, $\mbox{\boldmath $k$}_{\mbox{\scriptsize s}}$,
and $\mbox{\boldmath $G$}$ denote
a wave vector of the pump photon,
a wave vector of the signal photon,
and an arbitrary reciprocal lattice vector, respectively.
Because we consider a one-dimensional photonic crystal,
$\mbox{\boldmath $G$}$ is equal to zero or perpendicular to the long axis of the photonic crystal.
If we set $\mbox{\boldmath $G$}=\mbox{\boldmath $0$}$,
we can let $\mbox{\boldmath $k$}_{\mbox{\scriptsize p}}$ and $\mbox{\boldmath $k$}_{\mbox{\scriptsize s}}$ be parallel.

We consider the application of our method to the BB84 quantum key distribution protocol in the latter part of our study.
In reference~\cite{Adachi2007}, the utilization of the pairs of entangled photons generated by SPDC is proposed for the BB84 protocol.
In this scheme, Alice allows a nonlinear crystal to emit entangled photons flying in different directions with the SPDC.
Then, Alice detects one photon as the heralding signal and sends another photon to Bob.
Next, on the condition of Alice's detection,
Bob receives a single photon for the protocol.

As mentioned above, SPDC offers a superior source of single photons for the BB84 protocol.
The only limitation of the SPDC is that the conversion efficiency is very low.
In contrast, our method yields entangled pairs of photons $10^{5}$ times as efficiently as that with SPDC, and therefore, our proposal is very practical.

Some researchers provide a single-photon source and adopt a weak coherent light for which the average number of photons is
less than that for the signal of the BB84 protocol.
If we execute the BB84 protocol with a weak coherent light,
we encounter the following two technical problems:
\begin{enumerate}
\item Intervals of times for the emission of successive photons become random.
\item The number of photons included in a single pulse is random.
\end{enumerate}

The above two problems afford Eve clues for eavesdropping.
Alice and Bob cannot determine the time of the pulse emission because of the first problem.
Thus, Eve can steal coherent pulses in the middle of the quantum channel between legitimate users.
Moreover, Eve can inject false photons into the quantum channel.
In these cases, Alice and Bob cannot notice the malicious act of Eve.
If two photons are sent through the quantum channel by accident because of the second problem, Eve can place a beam splitter in the middle of the quantum channel,
steal one photon, and let the other one be untouched and reach Bob.
Alice and Bob can hardly detect the malicious act of Eve.

In reference~\cite{Hwang2003}, a decoy-state quantum key distribution protocol was proposed to detect the beam-splitting attack initiated by Eve.
In this method, Alice sends two coherent light pulses at random whose average number of photons is different from each other.
If Eve performs the attack with a beam splitter,
the difference between the average number of the photons of pulses detected by Bob becomes larger than that between the initial pulses,
and therefore, legitimate users become aware of the malicious acts of Eve.
The decoy-state method is considered an effective strategy against Eve in the BB84 protocol.

However, the advantage of the decoy-state method will be impaired considerably if Eve can execute the quantum nondemolition measurement to count the number of photons included in a single pulse.
The on-demand single-photon gun is ideal as the source of photons used in the BB84 protocol; coherent light is only a temporary alternative.

In contrast, entangled photons resolve the above two problems.
If Alice detects one photon with her detector as the heralding signal,
she informs Bob about it via the classical channel.
In this case, the probability that Bob receives a single photon is larger than that for the light of disentangled photons.
Hence, Alice and Bob can implement the BB84 protocol safely.

In this study, we consider LiNbO${}_{3}$ as a material with second-order nonlinear optical susceptibility $\chi^{(2)}$
because it is a typical medium that has large $\chi^{(2)}$.
Recently, experiments for fabricating photonic crystals comprised of LiNbO${}_{3}$ were executed eagerly \cite{Ishikawa2008b,Liang2017,Li2018,Li2020,Jiang2020}.
Technologies for manufacturing the samples of nonlinear photonic crystals were extremely refined in recent years.

In this paper, we assume that we can fabricate a one-dimensional photonic crystal constructed from layers of air and LiNbO${}_{3}$.
In fact, such photonic crystals are produced by plasma etching approaches
(for example, argon-ion milling) and electron-beam lithography in the laboratories
\cite{Liang2017,Li2018,Li2020,Jiang2020}.

One of the most remarkable quantities of this study is as follows.
After feeding a coherent light into the photonic crystal and beam splitter,
we obtain a state given by figure~\ref{Figure-04} from a port of the beam splitter on condition that we detect a single photon from the other port.
For this state,
probabilities of observing an odd number of photons are dominant and that of observing an even number of photons are suppressed.
Thus, it causes the bunching of an odd number of photons.
This state is neither a coherent state nor a squeezed state.
Hence, our result is novel from this point of view.

This paper is organized as follows.
In section~\ref{section-review-photonic-crystal},
we review how to generate the squeezed light with the photonic crystal.
In section~\ref{section-review-beam-splitter-squeezed-light},
we review how to convert the squeezed light into two entangled beams of photons.
In section~\ref{section-photon-counting-detection-entanglement},
we discuss how to detect entanglement with photon detectors.
In section~\ref{section-numerical-calculations-photon-statistics},
we perform the numerical calculations of photon statistics that reveals entanglement.
In section~\ref{section-photonic-crystal-parameters},
we calculate the physical parameters of the photonic crystal that generated the squeezed light efficiently.
In section~\ref{section-BB84},
we discuss the application of the entangled light generated by our method for the BB84 protocol.
In section~\ref{section-photonic-crystal-parameters-squeezing},
we study relationships between the squeezing parameter and the probabilities of detecting single photons.
In section~\ref{section-Realistic-setups},
we examine performances of realistic setups for the proposed method.
In section~\ref{section-Discussion},
we provide discussions.
In \ref{section-derivation-equations},
we explain details of derivations of some equations.

\section{\label{section-review-photonic-crystal}Generation of squeezed light with a nonlinear photonic crystal}
A brief review of reference~\cite{Sakoda2005} is provided in this section.

\begin{figure}
\centering
\includegraphics[width=8.6cm]{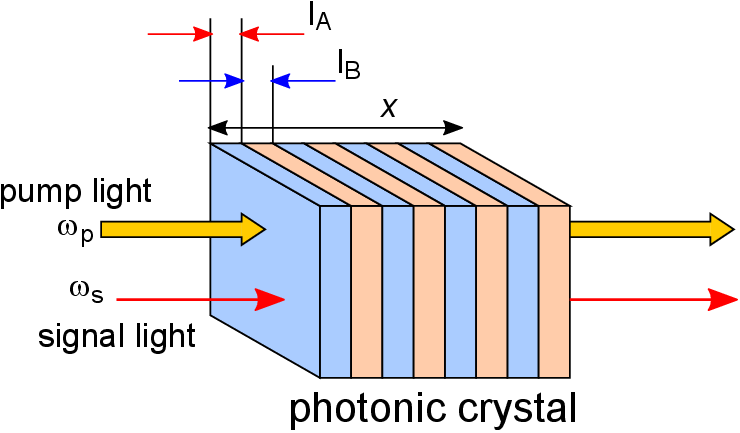}
\caption{\label{Figure-01}Schematic of nonlinear photonic crystal, and incident pump and signal light beams.}
\end{figure}

As indicated in figure~\ref{Figure-01},
we set the incident pump and signal light beams with angular frequencies $\omega_{\mbox{\scriptsize p}}$ and $\omega_{\mbox{\scriptsize s}}$,
respectively, into a nonlinear photonic crystal.
We assume that the incident light beams propagate in the $x$ direction, and their electric fields have only the $y$ component.
Further, we assume that the material with which the photonic crystal is constructed has a second-order nonlinear optical susceptibility $\chi^{(2)}$.
In concrete terms,
as shown in figure~\ref{Figure-01},
the photonic crystal is composed of two different materials in a one-dimensional stack
that have different linear refractive indices with one of the materials having a non-zero second-order nonlinearity.
For the sake of simplicity, 
we suppose that $\chi^{(2)}$ is real and positive.

We assume that the electric field of the pump light $E_{\mbox{\scriptsize p}}$ is sufficiently strong to allow us to neglect its quantum fluctuation.
Hence, we consider a classical pump light.
In contrast, the intensity of the signal light is very weak,
the number of photons of the light is small,
and we need to consider it as a quantum one.
Further, we assume that
$\omega_{\mbox{\scriptsize p}}
=
2\omega_{\mbox{\scriptsize s}}$
holds for angular frequencies.
We assume that wave vectors of the pump and signal light beams
$\mbox{\boldmath $k$}_{\mbox{\scriptsize p}}=(k_{\mbox{\scriptsize p}},0,0)$ and
$\mbox{\boldmath $k$}_{\mbox{\scriptsize s}}=(k_{\mbox{\scriptsize s}},0,0)$
satisfy
$k_{\mbox{\scriptsize p}}
=
2k_{\mbox{\scriptsize s}}$.

We can express the pump and signal light beams injected into the photonic crystal as
$\mbox{\boldmath $E$}_{\mbox{\scriptsize p}}
=
(0,E_{\mbox{\scriptsize p}},0)$,
$\hat{\mbox{\boldmath $E$}}_{\mbox{\scriptsize s}}
=
(0,\hat{E}_{\mbox{\scriptsize s}},0)$,
\begin{eqnarray}
E_{\mbox{\scriptsize p}}(x,t)
&=&
iA
\{
\exp[i(k_{\mbox{\scriptsize p}}x-\omega_{\mbox{\scriptsize p}}t+\theta)] \nonumber \\
&&
-
\exp[-i(k_{\mbox{\scriptsize p}}x-\omega_{\mbox{\scriptsize p}}t+\theta)]
\},
\end{eqnarray}
\begin{eqnarray}
\hat{E}_{\mbox{\scriptsize s}}(x,t)
&=&
i
\sqrt{\frac{\hbar\omega_{\mbox{\scriptsize s}}}{2\varepsilon_{0}V}}
\{
\hat{a}_{\mbox{\scriptsize s}}\exp[i(k_{\mbox{\scriptsize s}}x-\omega_{\mbox{\scriptsize s}}t)] \nonumber \\
&&
-
\hat{a}_{\mbox{\scriptsize s}}^{\dagger}\exp[-i(k_{\mbox{\scriptsize s}}x-\omega_{\mbox{\scriptsize s}}t)]
\},
\label{electric-field-incident-signal-light-0}
\end{eqnarray}
where $\theta$, $V$, and $A$ represent the phase of the pump light,
volume for quantization of the signal light, and amplitude of the pump light, respectively.
In equation~(\ref{electric-field-incident-signal-light-0}),
$\varepsilon_{0}$ denotes vacuum permittivity,
$\hbar$ is equal to $h/(2\pi)$,
and $h$ represents the Planck constant.

According to reference~\cite{Sakoda2005},
signal light that has passed the photonic crystal of width $x$ is described by the form
\begin{eqnarray}
\hat{E}(x,t)
&=&
i
\sqrt{\frac{\hbar\omega_{\mbox{\scriptsize s}}}{2\varepsilon_{0}V}}
\{
\hat{b}_{\mbox{\scriptsize s}}(x)\exp[i(k_{\mbox{\scriptsize s}}x-\omega_{\mbox{\scriptsize s}}t)] \nonumber \\
&&
-
\hat{b}_{\mbox{\scriptsize s}}^{\dagger}(x)\exp[-i(k_{\mbox{\scriptsize s}}x-\omega_{\mbox{\scriptsize s}}t)]
\},
\end{eqnarray}
\begin{equation}
\hat{b}_{\mbox{\scriptsize s}}(x)
=
\hat{a}_{\mbox{\scriptsize s}}\cosh(|\beta| x)
+
\hat{a}_{\mbox{\scriptsize s}}^{\dagger}\exp[i(\theta+\phi)]\sinh(|\beta| x),
\label{Bogoliubov-operator-0}
\end{equation}
where $\beta$ is given by
\begin{equation}
\beta
=
\frac{\omega_{\mbox{\scriptsize s}}A\chi^{(2)}}{\varepsilon_{0}v_{\mbox{\scriptsize g}}},
\label{photonic-crystal-squeezing-para-01}
\end{equation}
$\phi$ is given by
$e^{i\phi}=\beta/|\beta|$,
and
$v_{\mbox{\scriptsize g}}$ denotes the group velocity of the signal light in the photonic crystal.
An operator $\hat{b}_{\mbox{\scriptsize s}}(x)$ is obtained by the Bogoliubov transform of 
$\hat{a}_{\mbox{\scriptsize s}}$.

Here, we introduce a parameter
\begin{equation}
\zeta
=
\beta x \exp(i\theta)
=
|\beta|x \exp[i(\theta+\phi)],
\end{equation}
and define the squeezing operator
\begin{equation}
\hat{S}(\zeta)
=
\exp(
-\frac{\zeta}{2}\hat{a}_{\mbox{\scriptsize s}}^{\dagger 2}
+\frac{\zeta^{*}}{2}\hat{a}_{\mbox{\scriptsize s}}^{2}
).
\label{single-mode-squeezing-operator-01}
\end{equation}
Then, the following relation holds:
\begin{equation}
\hat{S}(\zeta)\hat{a}_{\mbox{\scriptsize s}}\hat{S}^{\dagger}(\zeta)
=
\hat{b}_{\mbox{\scriptsize s}}(x).
\end{equation}

If we place a coherent state $|\alpha\rangle$ as the signal light into the photonic crystal,
it emits a squeezed state $|-\zeta,\alpha\rangle$, where
\begin{eqnarray}
|-\zeta,\alpha\rangle
&=&
\hat{S}^{\dagger}(\zeta)|\alpha\rangle \nonumber \\
&=&
\hat{S}(-\zeta)|\alpha\rangle.
\end{eqnarray}

\section{\label{section-review-beam-splitter-squeezed-light}How to generate entangled light beams by putting the
squeezed light into a beam splitter}
A brief review of reference~\cite{Kim2002} is provided in this section. 

\begin{figure}[htp]
\centering
\includegraphics[width=8.6cm]{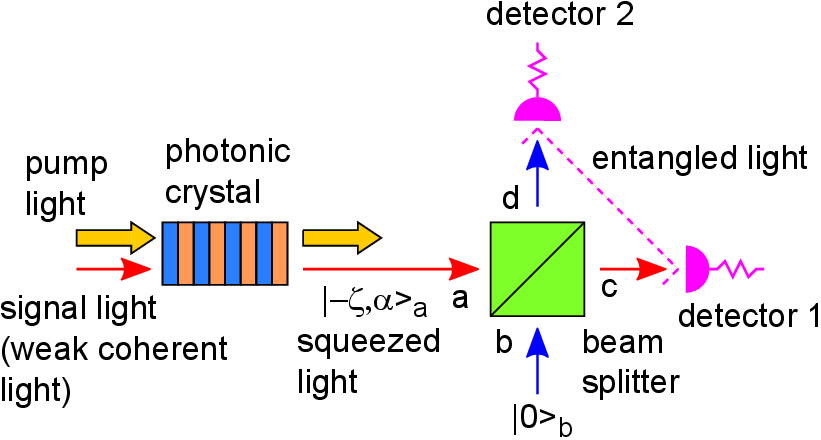}
\caption{\label{Figure-02}Schematic of method for generating entangled light beams with the photonic crystal and beam splitter
and method to detect entanglement by photon counting.
The squeezed light $|-\zeta,\alpha\rangle_{\mbox{\scriptsize a}}$ is injected into port $a$ of the beam splitter.
Because we do not consider port $b$, we describe its state as $|0\rangle_{\mbox{\scriptsize b}}$.}
\end{figure}

Figure~\ref{Figure-02} shows a schematic of the injection of the squeezed light into an input port $a$ of the beam splitter;
it is expressed as $|-\zeta,\alpha\rangle_{a}$.
We do not consider the other input port $b$, and we describe it as $|0\rangle_{b}$.
We let $\hat{a}$, $\hat{b}$, $\hat{c}$, and $\hat{d}$ denote the annihilation operators of ports $a$, $b$, $c$, and $d$, respectively.
[To avoid misunderstanding,
we draw attention to the fact that the operator $\hat{b}$ is different from the Bogoliubov transformed operator
$\hat{b}_{\mbox{\scriptsize s}}(x)$ given by equation~(\ref{Bogoliubov-operator-0}).]
The 50--50 beam splitter causes the following transformation:
\begin{eqnarray}
\hat{c}
&=&
\hat{B}(\delta)\hat{a}\hat{B}^{\dagger}(\delta), \nonumber \\
\hat{d}
&=&
\hat{B}(\delta)\hat{b}\hat{B}^{\dagger}(\delta),
\end{eqnarray}
\begin{equation}
\hat{B}(\delta)
=
\exp
[
\frac{\pi}{4}
(
\hat{a}^{\dagger}\hat{b}e^{i\delta}
-
\hat{a}\hat{b}^{\dagger}e^{-i\delta}
)
].
\label{transformation-B-delta-0}
\end{equation}

We examine the state of the light beams emitted from the ports of the beam splitter $c$ and $d$
if we feed
$|-\zeta,\alpha\rangle_{a}$ and $|0\rangle_{b}$
into ports $a$ and $b$, respectively.
Thus, we compute $\hat{B}^{\dagger}(\delta)|-\zeta,\alpha\rangle_{a}|0\rangle_{b}$.

The light emitted by the beam splitter is given by
\begin{equation}
\hat{B}^{\dagger}(\delta)|-\zeta,\alpha\rangle_{a}|0\rangle_{b}
=
\hat{B}^{\dagger}(\delta)
\hat{S}_{a}(-\zeta)
\hat{D}_{a}(\alpha)
|0\rangle_{a}|0\rangle_{b},
\end{equation}
where
\begin{eqnarray}
\hat{S}_{a}(-\zeta)
&=&
\exp(\frac{\zeta}{2}\hat{a}^{\dagger 2}-\frac{\zeta^{*}}{2}\hat{a}^{2}), \nonumber \\
\hat{D}_{a}(\alpha)
&=&
\exp(\alpha\hat{a}^{\dagger}-\alpha^{*}\hat{a}).
\label{Sa-Da-definition-0}
\end{eqnarray}

From slightly tough calculations, we obtain
\begin{eqnarray}
&&
\hat{B}^{\dagger}(\delta)
\hat{S}_{a}(-\zeta)
\hat{B}(\delta) \nonumber \\
&=&
\hat{S}_{ab}(-\frac{1}{2}\zeta e^{-i\delta})
\hat{S}_{a}(-\frac{1}{2}\zeta)
\hat{S}_{b}(-\frac{1}{2}\zeta e^{-2i\delta}),
\label{Bdagger-Sa-B-0}
\end{eqnarray}
where
\begin{equation}
\hat{S}_{ab}(\zeta)
=
\exp(-\zeta\hat{a}^{\dagger}\hat{b}^{\dagger}+\zeta^{*}\hat{b}\hat{a}).
\end{equation}
The details of the derivation of equation~(\ref{Bdagger-Sa-B-0}) are explained in \ref{subsection_Bdagger-Sa-B-0}.
Simultaneously, we obtain
\begin{equation}
\hat{B}^{\dagger}(\delta)
\hat{D}_{a}(\alpha)
\hat{B}(\delta)
=
\hat{D}_{a}(\frac{\alpha}{\sqrt{2}})
\hat{D}_{b}(\frac{\alpha}{\sqrt{2}}e^{-i\delta}),
\label{Bdagger-Da-B-0}
\end{equation}
\begin{equation}
\hat{B}^{\dagger}(\delta)|0\rangle_{a}|0\rangle_{b}
=
|0\rangle_{a}|0\rangle_{b},
\end{equation}
where
\begin{equation}
\hat{D}_{b}(\alpha)
=
\exp(\alpha\hat{b}^{\dagger}-\alpha^{*}\hat{b}).
\end{equation}
The details of the derivation of equation~(\ref{Bdagger-Da-B-0}) are explained in \ref{subsection_Bdagger-Da-B-0}.

Based on the above relations, we obtain 
\begin{eqnarray}
&&
\hat{B}^{\dagger}(\delta)|-\zeta,\alpha\rangle_{a}|0\rangle_{b} \nonumber \\
&=&
\hat{S}_{ab}(-\frac{1}{2}\zeta e^{-i\delta})
\hat{S}_{a}(-\frac{1}{2}\zeta)
\hat{S}_{b}(-\frac{1}{2}\zeta e^{-2i\delta}) \nonumber \\
&&
\times
\hat{D}_{a}(\frac{1}{\sqrt{2}}\alpha)
\hat{D}_{b}(\frac{1}{\sqrt{2}}\alpha e^{-i\delta})
|0\rangle_{a}|0\rangle_{b},
\label{beam-splitter-output-01}
\end{eqnarray}
where
\begin{equation}
\hat{S}_{b}(-\zeta)
=
\exp(\frac{\zeta}{2}\hat{b}^{\dagger 2}-\frac{\zeta^{*}}{2}\hat{b}^{2}),
\end{equation}
and modes $a$ and $b$ on the right-hand side of the equation correspond to ports $c$ and $d$, respectively.

A squeezing parameter that characterizes the light beam of port $a$ is given by $\zeta=|\beta| x\exp[i(\theta+\phi)]$,
where $\beta$ is defined in equation~(\ref{photonic-crystal-squeezing-para-01}).
Here, to simplify the discussion,
we set $\theta = \phi = 0$. 
Thus, from now on,
we write the squeezing parameter as $\zeta=|\beta| x=r$.
Further, for simplicity,
we set $\delta = 0$.
Then, the state of photons emitted from ports $c$ and $d$ is written in the form
\begin{eqnarray}
&&
\hat{B}^{\dagger}(0)|-r,\alpha\rangle_{a}|0\rangle_{b} \nonumber \\
&=&
\hat{S}_{ab}(-\frac{1}{2}r)
\hat{S}_{a}(-\frac{1}{2}r)
\hat{S}_{b}(-\frac{1}{2}r) \nonumber \\
&&
\times
\hat{D}_{a}(\frac{1}{\sqrt{2}}\alpha)
\hat{D}_{b}(\frac{1}{\sqrt{2}}\alpha)
|0\rangle_{a}|0\rangle_{b}.
\label{beam-splitter-output-entangled-01}
\end{eqnarray}

We draw attention to the fact that the operator $\hat{S}_{ab}(-r/2)$ appears in equation~(\ref{beam-splitter-output-entangled-01}).
This implies that photons emitted from ports $c$ and $d$ are entangled.
Hence, we obtain an entangled state by inserting squeezed light into the beam splitter.

\section{\label{section-photon-counting-detection-entanglement}Detection of the entanglement by photon counting}
We present our novel results in this section.

As shown in figure~\ref{Figure-02}, 
we observe two light beams generated by the photonic crystal and beam splitter using two detectors.
Each detector counts the number of photons in the pulse.
We expect to find a correlation between the number of photons counted by the two detectors if two light beams emitted from the ports of $c$ and $d$ by the beam splitter are entangled.

We let $n_{1}$ and $n_{2}$ denote the number of photons emitted from ports $c$ and $d$, respectively.
The probability that we obtain $n_{1}$ and $n_{2}$ is given as 
\begin{equation}
P(n_{1},n_{2})
=
|\langle n_{1},n_{2}|\hat{B}^{\dagger}(0)|-r,\alpha\rangle_{a}|0\rangle_{b}|^{2},
\label{P_n1_n2_formula-01}
\end{equation}
where $\hat{B}^{\dagger}(0)|-r,\alpha\rangle_{a}|0\rangle_{b}$
is given by equation~(\ref{beam-splitter-output-entangled-01}).

To compute this probability amplitude, we prepare some formulae \cite{Kim1989,deOliveira1990,Dantas1998},
\begin{equation}
\hat{S}_{a}(-\frac{1}{2}r)
\hat{D}_{a}(\frac{\alpha}{\sqrt{2}})
|0\rangle_{a}
=
\sum_{n = 0}^{\infty}
C_{n}^{a}|n\rangle_{a},
\label{Sa-Da-0state}
\end{equation}
\begin{equation}
C_{n}^{a}
=
\sum_{m = 0}^{\infty}
S_{nm}^{a}
D_{m}^{a},
\label{Cna-0}
\end{equation}
\begin{eqnarray}
&&
S_{nm}^{a} \nonumber \\
&=&
{}_{a}\langle n|\hat{S}_{b}(-\frac{1}{2}r)|m\rangle_{a} \nonumber \\
&=&
\sum_{l = 0}^{\lfloor n/2 \rfloor}
\sum_{k = 0}^{\lfloor m/2 \rfloor}
\delta_{n-2l,m-2k}
\frac{1}{l!k!}(-1)^{k}\frac{1}{2^{k+l}} \nonumber \\
&&
\times
\tanh^{k+l}(r/2)
\sqrt{\frac{m!n!}{(n-2l)!(m-2k)!}} \nonumber \\
&&
\times
\exp
\Biggl\{
-\frac{1}{2}[1+2(m-2k)]\ln[\cosh(r/2)]
\Biggr\},
\label{Snma-0}
\end{eqnarray}
\begin{eqnarray}
D_{m}^{a}
&=&
{}_{a}\langle m|\hat{D}_{a}(\frac{\alpha}{\sqrt{2}})|0\rangle_{a} \nonumber \\
&=&
\exp(-|\alpha|^{2}/4)\frac{1}{\sqrt{m!}}(\frac{\alpha}{\sqrt{2}})^{m}.
\label{Dma-0}
\end{eqnarray}
The details of the derivations of equations~(\ref{Cna-0}), (\ref{Snma-0}), and (\ref{Dma-0}) are explained in \ref{subsection_Cna-Snma-Dma}.
We can calculate coefficients $C_{n}^{b}$ in a similar manner.
The above calculations are gathered together into the following equations,
\begin{equation}
\langle n_{1},n_{2}|\hat{B}^{\dagger}(0)|-r,\alpha\rangle_{a}|0\rangle_{b}
=
\sum_{l = 0}^{\infty}
\sum_{k = 0}^{\infty}
S^{ab}_{n_{1}n_{2},lk}
C^{a}_{l}
C^{b}_{k},
\label{SabCaCb-formula-01}
\end{equation}
\begin{eqnarray}
S^{ab}_{n_{1}n_{2},lk}
&=&
\langle n_{1},n_{2}|
\hat{S}_{ab}(-\frac{1}{2}r)
|l\rangle_{a}|k\rangle_{b} \nonumber \\
&=&
\sum_{n = 0}^{\mbox{\scriptsize min}[n_{1},n_{2}]}
\sum_{m = 0}^{\mbox{\scriptsize min}[l,k]}
\delta_{l-m,n_{1}-n}
\delta_{k-m,n_{2}-n}
(-1)^{m} \nonumber \\
&&
\times
\tanh^{m+n}(r/2)
\frac{1}{m!n!} \nonumber \\
&&
\times
\exp
\Biggl\{
-(l+k-2m+1)\ln[\cosh(r/2)]
\Biggr\} \nonumber \\
&&
\times
\frac{\sqrt{l!k!n_{1}!n_{2}!}}{(l-m)!(k-m)!},
\label{Sn1n2lkab-01}
\end{eqnarray}
where details of the derivation of equation~(\ref{Sn1n2lkab-01}) are explained in \ref{subsection_Sn1n2lkab}.

\section{\label{section-numerical-calculations-photon-statistics}Numerical calculations of photon statistics that reveals entanglement}
We present the numerical calculations of $P(n_{1},n_{2})$ given by equations~(\ref{P_n1_n2_formula-01}) and (\ref{SabCaCb-formula-01}), respectively, in this section.

\begin{figure}
\centering
\includegraphics[width=8.6cm]{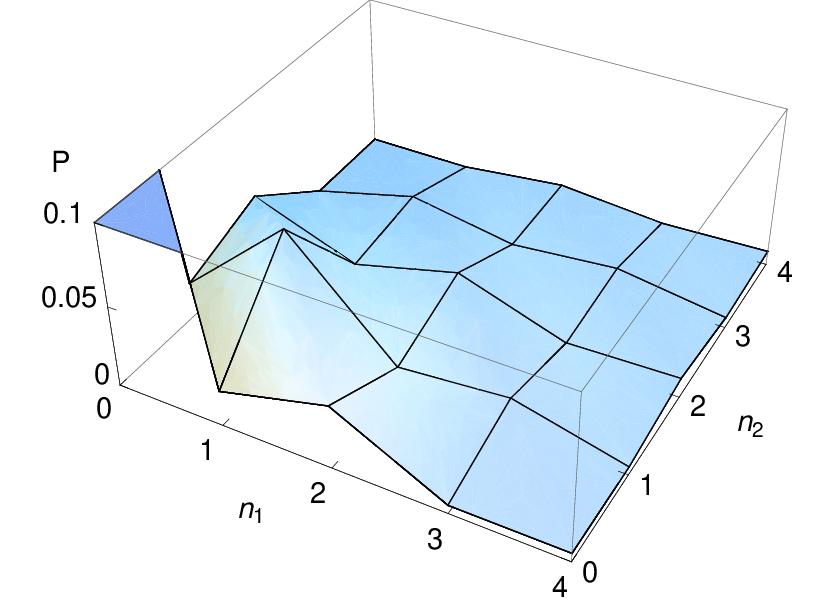}
\caption{\label{Figure-03}Graph of $P(n_{1},n_{2})$ for $\alpha = 1/2$ and $r = 1$.
The highest peak is $P(1,1)$ except for $P(0,0)$.}
\end{figure}

Figure~\ref{Figure-03} represents a two-dimensional graph of the probability $P(n_{1},n_{2})$
for the numbers of photons $n_{1}$ and $n_{2}$ obtained by detections in the ports $c$ and $d$, respectively, 
with $\alpha = 1/2$ and $r = 1$.
In figure~\ref{Figure-03}, we omit the point $P(0,0) = 0.417$ from the graph.
Looking at figure~\ref{Figure-03}, we note that $P(1,1)$ is the highest peak, except for $P(0,0)$.

\begin{figure}
\centering
\includegraphics[width=8.6cm]{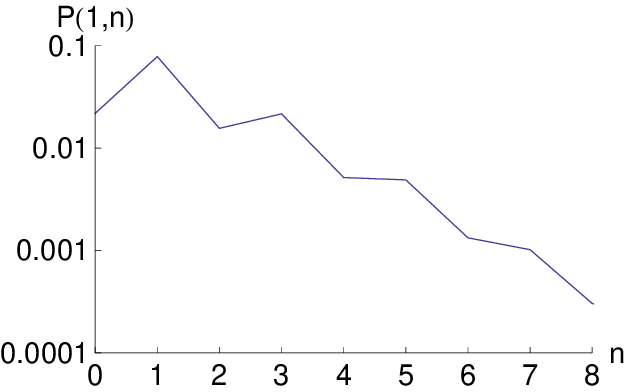}
\caption{\label{Figure-04}Graph $P(1,n)$ as a function of $n$ for $\alpha = 1/2$ and $r = 1$.
The vertical axis is displayed by the logarithmic scale. 
The probability $P(1,n)$ is dominant at odd $n$ and suppressed at even $n$.}
\end{figure}

We show another graph in figure~\ref{Figure-04} to examine this fact in detail.
Figure~\ref{Figure-04} shows a graph of $P(1,n)$ as a function of $n$ with setting $n_{1} = 1$.
The vertical axis of figure~\ref{Figure-04} is displayed in the logarithmic scale.

As shown in figure~\ref{Figure-04}, we become aware that $P(1,n)$ is promoted for odd $n$ and suppressed for even $n$.
The fact that the probability is dominant for odd $n$ is a reflection of quantum entanglement.
We can understand it from the following consideration.
Let us think about the single-mode squeezed vacuum state $\hat{S}(\zeta)|0\rangle$,
where the squeezing operator $\hat{S}(\zeta)$ is given by equation~(\ref{single-mode-squeezing-operator-01}).
In the squeezing operator $\hat{S}(\zeta)$,
the squeezed state $\hat{S}(\zeta)|0\rangle$ is a superposition of states whose number of photons is even because creation and annihilation operators appear in the forms $\hat{a}^{\dagger 2}$ and $\hat{a}^{2}$, respectively.
Hence, we can be sure that quantum entanglement makes the probabilities for odd $n$ dominant, as shown in figure~\ref{Figure-04}.

Here, we concentrate on events where the detector of port $c$ detects only a single photon. The probability that the detector of port $c$ detects only a single photon is described as

\begin{equation}
P_{1}
=
\sum_{n = 0}^{\infty}
P(1,n),
\label{definition-P1}
\end{equation}
and $P_{1} = 0.151$ for $\alpha = 1/2$ and $r = 1$.
Table~\ref{Table-01} shows $P(n)=P(1,n)/P_{1}$ for $n = 1, ..., 6$ that are renormalized probabilities.

\begin{table}
\centering
\caption{\label{Table-01}Table of $P(n)=P(1,n)/P_{1}$ for $\alpha = 1/2$ and $r = 1$.
The probability $P(1)$ is greater than $1/e = 0.368$,
which is the maximum probability of detecting only a single photon in a coherent light.}
\begin{tabular}{cccccccc}
\br
$n$ & $0$ & $1$ & $2$ & $3$ & $4$ & $5$ & $6$ \\ \mr
$P(n)$ & $0.145$ & $0.520$ & $0.104$ & $0.144$ & $0.0343$ & $0.0325$ & $0.008{\,}85$ \\
\br
\end{tabular}
\end{table}

Here, we focus on the following fact.
In table~\ref{Table-01}, we notice $P(1) = 0.520$.
In contrast, the probability that only a single photon is detected for a coherent state $|\beta\rangle$
($\beta$ is an arbitrary complex number) is given by $\exp(-|\beta|^{2})|\beta|^{2}$
and its maximum is equal to $1/e = 0.368$ for $|\beta|^{2} = 1$.
Thus, we cannot let the probability of detection of a single photon in the coherent state be larger than $0.368$.
Hence, the light beam emitted from port $d$ exceeds the limit of the coherent light because of the probability $P(1) = 0.520$.

In the above discussion, we assume that the squeezing parameter is equal to unity $r = 1$, 
for the squeezed light emitted from the photonic crystal.
It corresponds to $-4.34$ dB for the squeezing level.

\section{\label{section-photonic-crystal-parameters} Physical parameters of the photonic crystal for
generating a squeezed light}
We examine the physical parameters of the photonic crystal for generating a squeezed light by considering a concrete example.
We choose LiNbO${}_{3}$ as a typical material that has a large second-order nonlinear optical susceptibility $\chi^{(2)}$.

First, we investigate the conduction bands of photons for the photonic crystal.
The LiNbO${}_{3}$ is transparent for the light beam whose wavelength is between
$\lambda =  4 \times 10^{-7}$ m and $\lambda = 4.8\times 10^{-6}$ m
and its refractive index is given by $n\simeq 2.22$ \cite{Jundt1997,Gayer2008}.
For the sake of simplicity, we consider a one-dimensional photonic crystal.
As shown in figure~\ref{Figure-01}, we construct the photonic crystal by depositing the layers of materials A and B periodically with widths of $l_{\mbox{\scriptsize A}}$ and $l_{\mbox{\scriptsize B}}$, respectively.
We assume that materials A and B are air and LiNbO${}_{3}$, respectively.
The relative permittivity of air is given by $\varepsilon_{\mbox{\scriptsize A}}/\varepsilon_{0} = 1$.
We obtain the relative permittivity $\varepsilon_{\mbox{\scriptsize B}}/\varepsilon_{0}=n_{\mbox{\scriptsize B}}^{2}$ because the refractive index of LiNbO${}_{3}$ is given by $n_{\mbox{\scriptsize B}} = 2.22$.
As reported in references~\cite{Ishikawa2008b,Liang2017,Li2018,Li2020,Jiang2020},
we place $l_{\mbox{\scriptsize A}}=l_{\mbox{\scriptsize B}} = 5.5 \times 10^{-7}$ m.

\begin{figure}
\centering
\includegraphics[width=8.6cm]{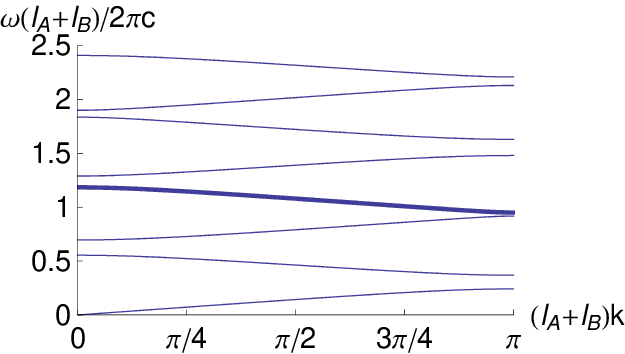}
\caption{\label{Figure-05}Graphs of the dispersion relation for the photonic crystal.
Each curve corresponds to a conductance band.
We show eight conduction bands from the bottom.
The thick curve represents the fourth conduction band from the bottom.
The band gaps imply that light waves of certain angular frequency ranges are not allowed to pass through the photonic crystal.}
\end{figure}

The dispersion relation $\omega=\omega(k)$ is given by describing the angular frequency of the incident light as $\omega$ and the wavenumber of photons in the photonic crystal as $k$ in the following equations as \cite{Azuma2008}
\begin{equation}
K_{\mbox{\scriptsize A}}
=
\frac{\omega}{c}\sqrt{\frac{\varepsilon_{\mbox{\scriptsize A}}}{\varepsilon_{0}}},
\label{dispersion-relation-1}
\quad
K_{\mbox{\scriptsize B}}
=
\frac{\omega}{c}\sqrt{\frac{\varepsilon_{\mbox{\scriptsize B}}}{\varepsilon_{0}}},
\end{equation}
\begin{eqnarray}
&&
\cos[(l_{\mbox{\scriptsize A}}+l_{\mbox{\scriptsize B}})k]
-
\cos(l_{\mbox{\scriptsize A}}K_{\mbox{\scriptsize A}})
\cos(l_{\mbox{\scriptsize B}}K_{\mbox{\scriptsize B}}) \nonumber \\
&&
+
\frac{K_{\mbox{\scriptsize A}}^{2}+K_{\mbox{\scriptsize B}}^{2}}{2K_{\mbox{\scriptsize A}}K_{\mbox{\scriptsize B}}}
\sin(l_{\mbox{\scriptsize A}}K_{\mbox{\scriptsize A}})
\sin(l_{\mbox{\scriptsize B}}K_{\mbox{\scriptsize B}})
=
0,
\label{dispersion-relation-2}
\end{eqnarray}
where $c$ denotes the speed of light in vacuum.
Figure~\ref{Figure-05} shows structures of the conduction bands obtained
by equations~(\ref{dispersion-relation-1}) and (\ref{dispersion-relation-2}).

\begin{figure}
\centering
\includegraphics[width=8.6cm]{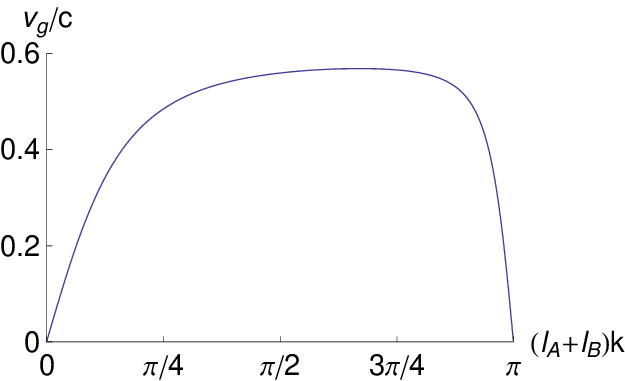}
\caption{\label{Figure-06}Group velocity of the fourth conduction band from the bottom.
At both ends of the graph, $v_{\mbox{\scriptsize g}}/c$ is equal to zero.}
\end{figure}

From the dispersion relation shown in figure~\ref{Figure-05},
we derive the group velocity of light in the photonic crystal.
If the wavenumber is equal to $k_{0}$,
the group velocity $v_{\mbox{\scriptsize g}}(k_{0})$ is given by
\begin{equation}
v_{\mbox{\scriptsize g}}(k_{0})
=
|
\frac{d\omega}{dk}(k_{0})
|.
\end{equation}
In figure~\ref{Figure-06}, we present a graph of the group velocity for the fourth conduction band from the bottom.
In figure~\ref{Figure-06}, the left end of the graph is represented by $k = 0$ and $v_{\mbox{\scriptsize g}} = 0$.
At this point, the angular frequency of the signal light is given by
$\omega_{\mbox{\scriptsize s}}(l_{\mbox{\scriptsize A}}+l_{\mbox{\scriptsize B}})/(2\pi c) = 1.18$.
This value corresponds to
the wavelength $\lambda_{\mbox{\scriptsize s}} = 2\pi c/\omega_{\mbox{\scriptsize s}} = 9.29\times 10^{-7}$ m.
This wavelength is included in the range of that of transparent light for LiNbO${}_{3}$.

Considering a double angular frequency of the signal light $\omega_{\mbox{\scriptsize s}}$,
{\it i.e.}, $\omega_{\mbox{\scriptsize p}} = 2\omega_{\mbox{\scriptsize s}}$,
we obtain $\omega_{\mbox{\scriptsize p}}(l_{\mbox{\scriptsize A}}+l_{\mbox{\scriptsize B}})/(2\pi c) = 2.37$;
the pump light is included in the eighth conduction band from the bottom.
The pump light is within the range of the wavelength of transparent light for LiNbO${}_{3}$ because of $\lambda_{\mbox{\scriptsize p}}=\lambda_{\mbox{\scriptsize s}}/2 = 4.65\times 10^{-7}$ m.

Therefore, we can generate the squeezed light from the signal and pump light beams
with $\omega_{\mbox{\scriptsize s}}$ and $\omega_{\mbox{\scriptsize p}}$,
respectively.
If we let the angular frequency of the signal light be equal to $\omega_{\mbox{\scriptsize s}}$,
its group velocity is given by zero.
Therefore, we need to ensure that the angular frequency of the signal light is slightly smaller than $\omega_{\mbox{\scriptsize s}}$.

The second-order nonlinear optical susceptibility of LiNbO${}_{3}$ is given by
$\chi^{(2)}=\varepsilon_{0}\tilde{\chi}^{(2)}$ and
$\tilde{\chi}^{(2)} = 25.2\times10^{-12}$ m/V \cite{Shoji1997,Kawase2002,Schiek2012}.
Then, the squeezing parameter $\zeta$ of the photonic crystal is given as
\begin{equation}
\zeta=\beta l,
\label{zeta-formula}
\end{equation}
\begin{equation}
\beta
=
\frac{\omega_{\mbox{\scriptsize s}}A\tilde{\chi}^{(2)}}{v_{\mbox{\scriptsize g}}},
\label{beta-formula}
\end{equation}
where $l$ represents the total length of the layers of LiNbO${}_{3}$ in the photonic crystal,
and $l = 5.0\times 10^{-5}$ m.
In equation~(\ref{beta-formula}), $A$ denotes the amplitude of the electric field $E$, and its unit is V/m.
$v_{\mbox{\scriptsize g}}$ denotes the group velocity of the signal light, 
and $\omega_{\mbox{\scriptsize s}}$ represents an angular frequency of the signal light.

Here, we consider the following approximation.
Figures~\ref{Figure-05} and \ref{Figure-06} show that the group velocity of the signal light $v_{\mbox{\scriptsize g}}$ largely moves from zero to a positive value 
by changing the angular frequency in small amounts
from $\omega_{\mbox{\scriptsize s}}(l_{\mbox{\scriptsize A}}+l_{\mbox{\scriptsize B}})/(2\pi c) = 1.18$.
Hence, in equation~(\ref{beta-formula}), we fix $\omega_{\mbox{\scriptsize s}}$ at the left end of the conduction band
and consider $v_{\mbox{\scriptsize g}}$ as a variable.

Here, we let the squeezing parameter $\zeta$ be equal to unity; then, because $\zeta = 1$,
the amplitude $A$ is represented as a function of the group velocity $v_{\mbox{\scriptsize g}}$.
In concrete terms, from $\beta l = 1$, we obtain
\begin{equation}
A
=
\frac{v_{\mbox{\scriptsize g}}}
{
\omega_{\mbox{\scriptsize s}}
\tilde{\chi}^{(2)}
l
}
=
1.17\times 10^{8}\times \frac{v_{\mbox{\scriptsize g}}}{c}\quad\mbox{V/m}.
\label{A-formula}
\end{equation}

Here, we estimate $A$ practically.
The relation between the intensity $I$[W/m${}^{2}$] and amplitude $A$[V/m] of the electric field of the pump laser is given as
\begin{equation}
I=\frac{1}{2}\varepsilon_{0}c n A^{2},
\label{relation-I-A-0}
\end{equation}
where $n$ denotes the refractive index of vacuum.
The standard specification of a semiconductor laser for commercial use is given by the radiant flux $W = 0.03$ W
and the radius of the laser beam $d = 5.0\times 10^{-6}$ m.
Since equation~(\ref{relation-I-A-0}), 
\begin{equation}
I=\frac{W}{\pi d^{2}},
\label{I-W-relation-0}
\end{equation}
and the approximation $n = 1$,
we get $A = 5.36\times 10^{5}$ V/m.
Thus, we have 
\begin{equation}
\frac{v_{\mbox{\scriptsize g}}}{c}
=
4.57\times 10^{-3}.
\label{accuracy-group-velocity-0}
\end{equation}

Therefore, 
we have to adjust the group velocity of the light passing through the photonic crystal
with accurate numerical precision of order $10^{-3}$ to obtain the squeezing parameter $\zeta = 1$.
Hence, we must change the frequency of the signal light precisely such that its group velocity is provided by equation~(\ref{accuracy-group-velocity-0}).
In figure~\ref{Figure-06}, the variance of $v_{\mbox{\scriptsize g}}/c$ between zero and $4.57\times 10^{-3}$ corresponds to that of
$(l_{\mbox{\scriptsize A}}+l_{\mbox{\scriptsize B}})k$ between zero and $4.31\times 10^{-3}$ approximately.
This variance of $(l_{\mbox{\scriptsize A}}+l_{\mbox{\scriptsize B}})k$ corresponds to
$\Delta\omega = 1.96\times 10^{9}$ 1/s
in figure~\ref{Figure-05}.
Hence, we need to adjust the frequency of the signal light with a precision of approximately $\Delta\nu = \Delta\omega/(2\pi)=3.11\times 10^{8}$ Hz.

In the above estimation,
the frequency of the signal light is given by
$\nu_{\mbox{\scriptsize s}}=\omega_{\mbox{\scriptsize s}}/(2\pi) = 3.23\times 10^{14}$ Hz,
and it is slightly higher than that of the terahertz radiation.

\section{\label{section-BB84}Application to the BB84 protocol}
We can utilize our method of generating entangled light beams as the source of photons for the BB84 protocol. In this section, we discuss this application.

\begin{figure}
\centering
\includegraphics[width=8.6cm]{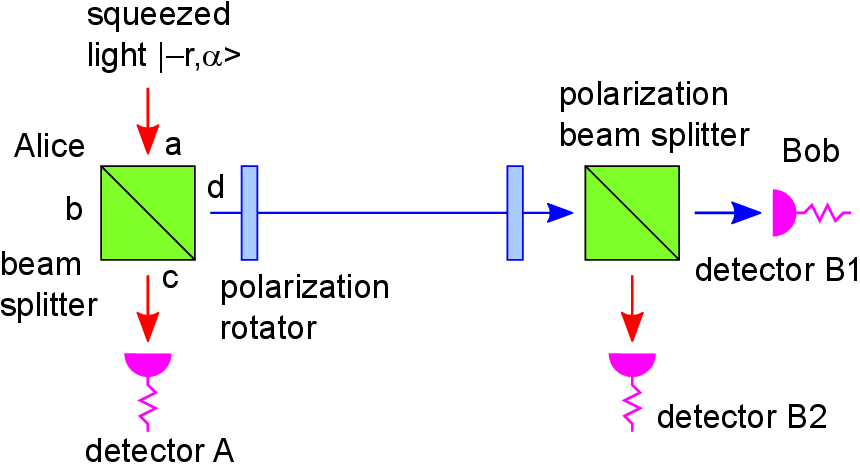}
\caption{\label{Figure-07}Schematic of the application of the entangled light beams emitted from the beam splitter to the BB84 protocol.
Both Alice and Bob use polarization rotators for choosing the basis at random.}
\end{figure}

Figure~\ref{Figure-07} shows a schematic of the setup of an experiment for realizing the BB84 protocol with entangled light beams.
Alice places the squeezed light emitted from the photonic crystal into port $a$ of the beam splitter.
Alice detects photons flying out of port $c$ with detector A as the heralding signal.
Photons emitted from port $d$ are transformed into states $\{|\uparrow\rangle,|\downarrow\rangle,|+\rangle,|-\rangle\}$
with the polarization rotator, and they are sent to Bob.
Bob receives these photons, chooses the basis for the measurement,
$\{|\uparrow\rangle,|\downarrow\rangle\}$ or
$\{|+\rangle,|-\rangle\}$,
splits them with the polarization beam splitter,
and detects them with detectors B1 and B2.

We consider a scenario where a probability amplitude for the state
$|1\rangle_{a}|1\rangle_{b}$
emerges by injecting the squeezed state $|-r,\alpha\rangle_{a}$ into the 50--50 beam splitter.
Then, the photon emitted from port $c$ is detected by detector A as the heralding signal.
From this event, Alice and Bob can confirm the time when the beam splitter emits the two photons.
Then, the other photon flies from the port $d$ to Bob.
This process is equivalent to the generation of an on-demand single photon.

If Alice and Bob co-operate using quantum entanglement,
they can detect Eve's malicious acts.
Their co-operation is given as follows.

Figure~\ref{Figure-04} shows the plot of $P(1,n)$ for $n = 0,1,...,8$, which are the probabilities of the condition that Alice detects only a single photon with detector A.
The probability that $n_{1} = 1$ and $n_{2} = 1$ hold is $0.0783$.
This case is equivalent to the process of the on-demand photon gun.
However, the probability for $n_{1} = 1$ and $n_{2} = 3$ is given by $P(1,3) = 0.0216$.
In this case, Eve can place a beam splitter in the middle of the quantum channel and steal a photon out of the three photons ($n_{2} = 3$).

We can avoid this issue using the following method.
Assume that Eve placed the 50--50 beam splitter halfway through the quantum channel.
If $(n_{1},n_{2})=(1,3)$ holds, the average number of photons Eve can steal is $1.5$.
In contrast, if $(n_{1},n_{2})=(1,1)$ holds, the following can occur.
Eve steals one photon and sends no photons to Bob with a probability $1/2$,
or Eve steals no photon and sends one photon to Bob with a probability $1/2$.
Therefore, probability that Bob detects no photon increases by $P(1,1)/2 = 0.0391$; this increment is detectable.
Therefore, if Bob monitors the probability of the detection of photons during the protocol, he can notice the malicious acts of Eve.

The scenario is slightly complicated when we consider an actual and practical setup.
With the current technology, Alice and Bob's detectors can often detect only a number of photons exceeding the voltage threshold because the realization of a single-photon detector is difficult.
Therefore, Alice can hardly distinguish an event of a single photon emitted from port $c$ among those with multiple photons.

We define the following three probabilities:
\begin{equation}
Q^{(1)}
=
\sum_{n_{1} = 1}^{\infty}
\sum_{n_{2} = 0}^{\infty}
P(n_{1},n_{2}),
\end{equation}
\begin{equation}
Q^{(2)}
=
\sum_{n_{1} = 1}^{\infty}
\sum_{n_{2} = 1}^{\infty}
P(n_{1},n_{2}),
\end{equation}
\begin{equation}
Q^{(3)}
=
\sum_{n_{1} = 1}^{\infty}
P(n_{1},1).
\end{equation}
The first one is
a probability that Alice detects photons with detector A.
The second one is
a probability that Alice detects photons with detector A
and the number of photons sent to Bob is equal to or greater than one.
The third one is
a probability that Alice detects photons with detector A and only one photon is sent to Bob.
In the current case,
they are given by
$Q^{(1)} = 0.509$,
$Q^{(2)} = 0.435$,
and $Q^{(3)} = 0.129$.

Thus, if Eve puts the 50--50 beam splitter in the middle of the quantum channel,
the probability that Bob does not detect photons changes
from $Q^{(1)}-Q^{(2)} = 0.074$ to $Q^{(1)}-Q^{(2)}+(Q^{(3)}/2) = 0.139$
under the condition that Alice detects the photons.
Alice and Bob can see through the malicious acts of Eve by monitoring the change in probability.

The above discussion relies on the assumption that Eve cannot count the number of photons of a pulse on a quantum channel by quantum nondemolition measurement or steal photons partially from a pulse that contains more than two photons.

\section{\label{section-photonic-crystal-parameters-squeezing} Relations between the squeezing parameter
and probabilities $P(1,1)$, $P_{1}$, and $P(1)$}
\begin{figure}
\centering
\includegraphics[width=8.6cm]{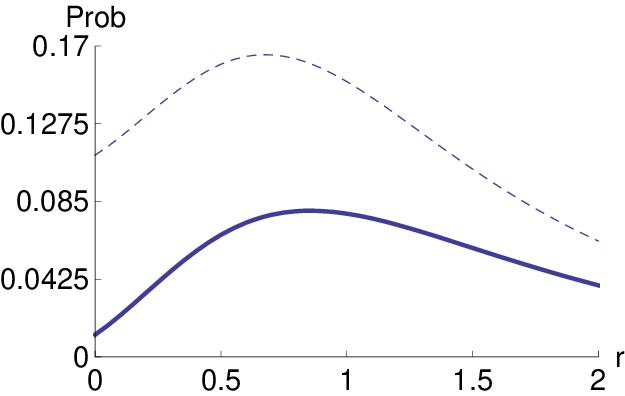}
\caption{\label{Figure-08}Graphs of $P(1,1)$ and $P_{1}$ as functions of $r$ for $\alpha = 1/2$.
The thick solid and thin dashed curves represent $P(1,1)$ and $P_{1}$, respectively.
The maximum value of $P(1,1)$ is given by $0.0799$ at $r = 0.85$; 
the maximum value of $P_{1}$ is equal to $0.165$ at $r = 0.675$.}
\end{figure}

\begin{figure}
\centering
\includegraphics[width=8.6cm]{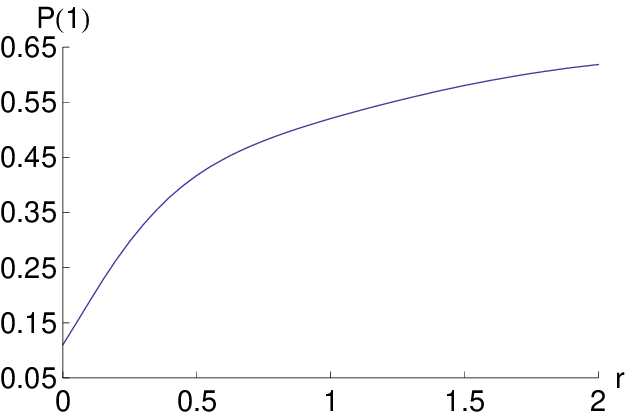}
\caption{\label{Figure-09}Graph of $P(1)$ as a function of $r$ for $\alpha = 1/2$.
$P(1)$ increases with an increase in $r$.}
\end{figure}

The squeezing parameter of the squeezed light generated by the photonic crystal $r=|\zeta|$ is defined
by equations~(\ref{zeta-formula}) and (\ref{beta-formula}).
We plotted probabilities $P(1,1)$ given by equation~(\ref{P_n1_n2_formula-01}) and $P_{1}$ defined by equation~(\ref{definition-P1}),
as functions of $r$ by setting $\alpha = 1/2$ in figure~\ref{Figure-08}.
The maximum values of $P(1,1)$ and $P_{1}$ are $0.0799$ at $r = 0.85$
and $0.165$ at $r = 0.675$, respectively.
Figure~\ref{Figure-09} shows a graph of $P(1)=P(1,1)/P_{1}$ against $r$ with $\alpha = 1/2$.
The value $P(1)$ implies the probability that a single photon is detected on port $d$ under the condition that the other single photon is detected on port $c$, and therefore, it can be a characteristic index of similarity between the on-demand single-photon source and our method.

The larger $r$ promises the realization of the photon gun and reveals the properties of quantum mechanics significantly because $P(1)$ becomes greater in proportion to the increase in $r$ in figure~\ref{Figure-09}.
We let $r$ be larger to apply our method to the single-photon source for the BB84 protocol.

\section{\label{section-Realistic-setups}Performances of realistic setups for the proposed method}
So far, we investigate the proposed method with optimum parameters that are sometimes very difficult to realize
in actual devices.
Thus, in this section,
we examine performances of realistic setups for the proposal,
for example,
with a more practical group velocity, with loss of propagating photons, and so on.

\subsection{\label{subsection-width-group-velocity}
A relationship between variations of the width of layers of the one-dimensional photonic crystal
and the accuracy of the group velocity of the signal photons}
\begin{figure}
\centering
\includegraphics[width=8.6cm]{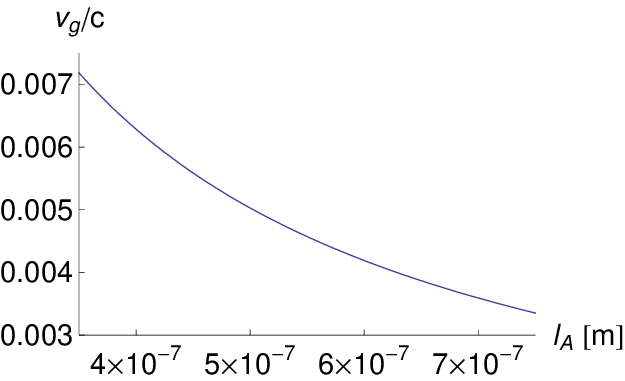}
\caption{\label{Figure-10}Graph of $v_{\mbox{\scriptsize g}}/c$ as a function of $l_{\mbox{\scriptsize A}}$.
The unit of $l_{\mbox{\scriptsize A}}$ is m
and $v_{\mbox{\scriptsize g}}/c$ is a dimensionless quantity.
The other physical parameters are given in section~\ref{section-photonic-crystal-parameters}.}
\end{figure}

In actual setups,
the width of layers of the one-dimensional photonic crystal may varies because of imperfect production processes.
Setting $l_{\mbox{\scriptsize A}}=l_{\mbox{\scriptsize B}}$,
we evaluate $v_{\mbox{\scriptsize g}}/c$ as a function of $l_{\mbox{\scriptsize A}}$ and plot it in figure~\ref{Figure-10}.
From figure~\ref{Figure-10},
we note that $v_{\mbox{\scriptsize g}}/c$ is in inverse proportion to $l_{\mbox{\scriptsize A}}$.
For example,
if $l_{\mbox{\scriptsize A}}=500$ nm, $550$ nm, and $600$ nm,
$v_{\mbox{\scriptsize g}}/c$ is given by $5.03\times 10^{-3}$, $4.57\times 10^{-3}$, and $4.19\times 10^{-3}$, respectively.
The slope of the curve is gentle in figure~\ref{Figure-10}.
Thus, we can conclude that the group velocity $v_{\mbox{\scriptsize g}}/c$ is stable with respect to
the fluctuation of the width of the layers of the photonic crystal $l_{\mbox{\scriptsize A}}$.

\subsection{\label{subsection-precisions-frequency}
Required precisions of the frequency of the signal light}
\begin{figure}
\centering
\includegraphics[width=8.6cm]{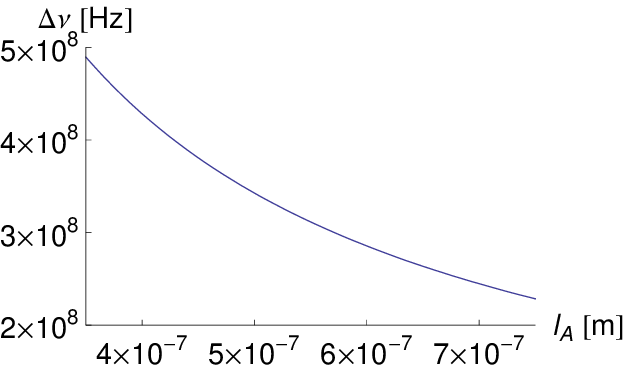}
\caption{\label{Figure-11}Graph of $\Delta\nu$,
the required precision of the frequency of the signal light to realize $v_{\mbox{\scriptsize g}}/c=4.57\times 10^{-3}$,
as a function of $l_{\mbox{\scriptsize A}}$.
The units of $\Delta\nu$ and $l_{\mbox{\scriptsize A}}$ are Hz and m, respectively.
The other physical parameters are given in section~\ref{section-photonic-crystal-parameters}.}
\end{figure}

To achieve $v_{\mbox{\scriptsize g}}/c\sim 4.57\times 10^{-3}$,
we must adjust the frequency of the signal light precisely
because $v_{\mbox{\scriptsize g}}/c$ is sensitive to a change of the wavenumber of the signal light as shown in figure~\ref{Figure-06}.
In figure~\ref{Figure-11},
we plot $\Delta\nu$, the required precision of the frequency of the signal light for attaining
$v_{\mbox{\scriptsize g}}/c=4.57\times 10^{-3}$,
as a function of $l_{\mbox{\scriptsize A}}$.
The typical $\Delta\nu$ is given by $3.11\times 10^{8}$ Hz for $l_{\mbox{\scriptsize A}}=5.5\times 10^{-7}$ m
and we must adjust $\nu$ with $\Delta\nu$ of precision
for an actual experiment.
Looking at figure~\ref{Figure-11},
we note that $\Delta\nu$ is in inverse proportion to $l_{\mbox{\scriptsize A}}$.
The slope of the curve is gentle in figure~\ref{Figure-11}.
Thus, although we need a high accuracy $\Delta\nu/\nu=9.65\times 10^{-7}$,
$\Delta\nu$ itself is stable with respect to the fluctuation of $l_{\mbox{\scriptsize A}}$.

\subsection{\label{subsection-probability-group-velocity}
A relationship between the probability that a single photon is detected and the group velocity}
\begin{figure}
\centering
\includegraphics[width=8.6cm]{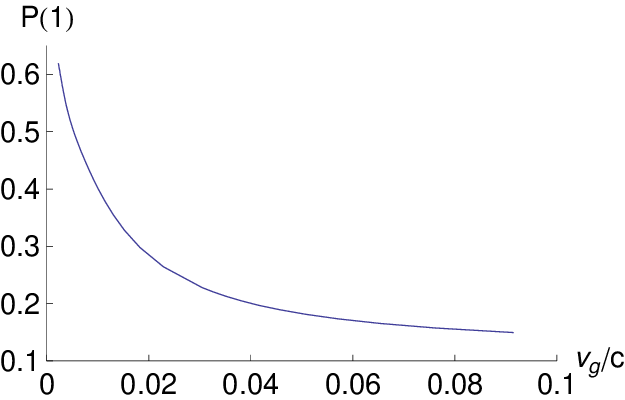}
\caption{\label{Figure-12}Graph of $P(1)$ as a function of $v_{\mbox{\scriptsize g}}/c$ for $\alpha=1/2$.
The other physical parameters are given in section~\ref{section-photonic-crystal-parameters}.}
\end{figure}

In figure~\ref{Figure-09},
we show the graph of $P(1)$,
the probability that the single photon is detected on the condition of the single heralding photon,
as a function of $r$ for $\alpha=1/2$.
We can redraw this graph as a function of $v_{\mbox{\scriptsize g}}/c$
because the squeezing parameter $r=|\zeta|$ is given by equations~(\ref{zeta-formula}) and (\ref{beta-formula}).
Figure~\ref{Figure-12},
which is essentially equivalent to figure~\ref{Figure-09},
shows a graph of $P(1)$ as a function of $v_{\mbox{\scriptsize g}}/c$.
Looking at figure~\ref{Figure-12},
we note that $P(1)$ decreases rapidly as $v_{\mbox{\scriptsize g}}/c$ becomes larger.
For example, when $v_{\mbox{\scriptsize g}}/c\geq 0.0122$,
$P(1)\leq 1/e=0.368$ holds and our method becomes useless as the single-photon source.

\subsection{\label{subsection-loss}
Loss of the propagating photons in the photonic crystal}
Unfortunately, a realistic photonic crystal is not transparent and we cannot neglect loss of the propagating photons.
In references~\cite{Olivier2003,Gerace2004,OFaolain2006},
it was reported that the propagation loss of the practical photonic crystals was
from $-10$dB/cm to $-105$dB/cm.
To obtain a large squeezing parameter $r=|\zeta|$,
we better let $l$, the total length of the one-dimensional photonic crystal, be longer.
However, as the total length $l$ gets longer,
loss of the signal photons becomes more serious and we cannot neglect it.

\begin{figure}
\centering
\includegraphics[width=8.6cm]{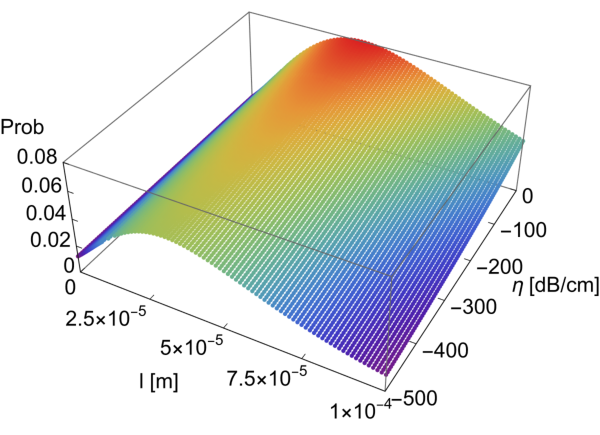}
\caption{\label{Figure-13}Graph of
$\mbox{Prob}(l,\eta)=P(1,1)\exp(-|\eta|l)$,
the probability of obtaining the single photon with the heralding photon,
as a function of the total length $l$ and the rate of the propagation loss $\eta$ for $\alpha=1/2$.
The units of $l$ and $\eta$ are m and dB/cm, respectively.
The other physical parameters are given in section~\ref{section-photonic-crystal-parameters}.}
\end{figure}

Thus, the squeezing parameter and the loss of propagating photons are in a trade-off relationship.
In figure~\ref{Figure-13}, we plot a graph of
$\mbox{Prob}(l,\eta)=P(1,1)\exp(-|\eta|l)$,
the probability of obtaining the single photon with the heralding photon,
as a function of the total length $l$ and the rate of the propagation loss $\eta$.
On one hand,
the squeezing parameter $r=|\zeta|$ depends on $l$ according to equations~(\ref{zeta-formula}) and (\ref{beta-formula}),
and $P(1,1)$ depends on $r$ according to equations~(\ref{beam-splitter-output-entangled-01}) and (\ref{P_n1_n2_formula-01}).
On the other hand,
$\mbox{Prob}(l,\eta)$ decreases exponentially as $l$ gets longer.

In figure~\ref{Figure-13}, $\mbox{Prob}(l,\eta)$ becomes maximum, that is, $\mbox{Prob}=0.0799$,
when $l$ and $\eta$ are given by $4.25\times 10^{-5}$ m and zero dB/cm, respectively.
However, if we set $\eta=-500$ dB/cm,
$\mbox{Prob}(l,\eta)$ is given by $0.0490$ for $l=4.25\times 10^{-5}$ m.
Setting $l=4.25\times 10^{-5}$ m,
$\mbox{Prob}(l,\eta)$ is equal to $0.0725$ for $\eta=-100$ dB/cm.
Thus, we can allow the loss of the propagating photons up to $-100$ dB/cm around.
This requirement is not severe for the current technology.

Because of $\alpha=1/2$, $\mbox{Prob}=0.0725$ implies that the conversion efficiency of the entangled pair is $0.290$
per signal photon for $\eta=-100$ dB/cm.
Thus, we can conclude that our method can keep a competitive advantage over the SPDC process of the PPLN
(the best conversion efficiency: $4\times 10^{-6}$ per pump photon)
\cite{Bock2016}
even if we consider the loss of photons.

\section{\label{section-Discussion}Discussion}
First of all, we make a rigorous comparison between the state-of-the-art alternative approaches and the proposed method.

Luo {\it et al} studied generation of path-entangled photons
from the SPDC process of two-dimensional quadratic nonlinear photonic crystals \cite{Luo2017}.
They used a two-dimensional hexagonally poled lithium tantalate (HPLT) and obtained the conversion efficiency $1.25\times 10^{-8}$
per pump photon.
Jin {\it et al} conducted experiments of the SPDC
in nonlinear plasmonic metasurface that is composed of periodic silver nanostripes placed on top of a homogeneous LiNbO${}_{3}$
\cite{Jin2021}.
They reached the conversion efficiency of $1.8\times 10^{-6}$ for generation of entangled photon pairs.
As mentioned in section~\ref{section-Introduction},
the best conversion efficency of the SPDC is $4\times 10^{-6}$ per pump photon
for the PPLN waveguide \cite{Bock2016}.

If we use the entangled photon pair state as a source of the single photon,
the heralding efficiency is important.
In our method, the heralding efficency attains $0.520$ as shown in table~\ref{Table-01}.
Zhong {\it et al} demonstrated $97${\%} heralding efficiency from an ultrafast pulsed laser pumped SPDC using BBO
although their main motivation of experiments was generation of 12-photon entanglement with 12 SPDC sources \cite{Zhong2018}.

Cui {\it et al} proposed how to achieve quantum state engineering by an $SU(1,1)$ nonlinear interferometer
with a pulsed pump and a controllable linear spectral phase shift \cite{Cui2020}.
Li {\it et al} demonstrated this method experimentally and reached a heralding efficiency of $90${\%}
at a brightness of $0.039$ photons/pulse \cite{J-Li2020}.

Next, we compare a recent result of the PPLN and the proposed method from the point of view of squeezing.
In reference~\cite{Eto2008}, Eto {\it et al} investigated the utilization of periodically poled MgO:LiNbO${}_{3}$ (PPLN) waveguide
for generating squeezed light.
They obtained a squeezing of $-4.1$ dB, which is almost the same as the squeezing level of our method.

In the experiment conducted in reference~\cite{Eto2008}, the radiant flux values for the pump light and signal light are set
to $3.7\times 10^{-4}$ W and $2.0\times 10^{-7}$ W, respectively.
The wavelength and time duration of an incident pulse for the signal light are given
by $\lambda = 1.535\times 10^{-6}$ m and $\tau = 3.7\times 10^{-9}$ s, respectively.
The radius of the beam of the signal light and refractive index of the air are equal to $d = 5.0\times 10^{-6}$ m and $n = 1$, respectively.
From 
equations~(\ref{relation-I-A-0}) and (\ref{I-W-relation-0}),
we obtain the amplitude of the electric field of the signal light as $A = 1.39\times 10^{3}$ V/m.

To convert the amplitude of the electric field $A$ into the number of photons $N$,
we use 
\begin{equation}
N
=
A^{2}
\frac{2\varepsilon_{0} V}{\hbar\omega},
\end{equation}
where $\omega$ and $V$ denote the angular frequency of the signal light and the volume for the quantization of photons, respectively.
We estimate the lengths of the three sides of $V$ at $c\tau = 1.11$ m, $d = 5.0\times 10^{-6}$ m, and $d = 5.0\times 10^{-6}$ m, respectively.
From these calculations, we obtain the number of photons as $N = 7.28\times 10^{3}$.

The experiment in reference~\cite{Eto2008} manipulated a considerably larger number of photons than the proposed method; therefore, the process described in reference~\cite{Eto2008} and our proposals are completely different from each other.
In our method, we draw attention to the fact that only a few photons are placed in the photonic crystal.
Thus, for letting coherent light whose average number of photons is less than the one to be squeezed,
our proposal is superior to the experiment of reference~\cite{Eto2008}.

The length of the PPLN waveguide used in reference~\cite{Eto2008} was approximately a few centimeters, and the authors reported that it is caused
loss of signal by $4$ dB ($39.8$ percent).
This fact is a serious disadvantage for utilizing the PPLN waveguide.
This loss of signal causes classical dissipation,
and therefore, the squeezed light passing through the PPLN waveguide tends to lack the properties of quantum mechanics.
We expect that the weak squeezed light whose number of photons is less than one is easily destroyed by such dissipation.

Our method uses a photonic crystal made of LiNbO${}_{3}$, whose length is approximately dozens of micrometers; therefore, photons travel a short distance and can avoid classical dissipation drastically.
Hence, we can obtain weak squeezed light without losing the properties of quantum mechanics using our method efficiently.

In section~\ref{section-BB84}, we assume that Alice and Bob can use detectors that can detect only a number of photons
exceeding the voltage threshold; they cannot count a few photons exactly in the BB84 protocol.
In reference~\cite{Yuan2007}, the development of a single-photon detector suitable for the BB84 protocol was reported.
This single-photon detector was originally designed for the BB84 protocol with a weak coherent light; however, this device can be used for our method.
In section~\ref{section-BB84}, we discussed the detection of the eavesdropper with a photon detector
for the voltage threshold.
When we utilize the single-photon detector,
we can omit this cumbersome process.

In this paper, we discussed the generation of entangled photon beams with a nonlinear photonic crystal and a beam splitter
and its application to the BB84 protocol.
The notablest advantage of our method is that its yield of producing entangled pairs of photons is $10^{5}$ times as efficient as that with conventional SPDC.
The limitation of our work is that we must adjust the frequency of the signal light fed to the photonic crystal precisely,
for example, $\Delta\nu/\nu_{\mbox{\scriptsize s}}=9.65\times 10^{-7}$.
Our results provide a superior single-photon source for various quantum communication processes,
particularly the BB84 protocol,
compared to a weak coherent light and SPDC.
When we apply our method to the BB84 protocol,
Bob monitors the probability that he does not detect photons to see through malicious acts of Eve.
We can regard this ingenuity as an improvement of the decoy-state method.

Some researchers might say that our method requires a photonic crystal where the group velocity is reduced to $4.57\times 10^{-3}$ times
the speed of light in vacuum from equation~(\ref{accuracy-group-velocity-0})
and it is an unrealistic value if disorder in the crystal is taken into account.
However, we can expect further progress in techniques of fabricating a photonic crystal made of LiNbO${}_{3}$.
We anticipated that our method will provide great advantages not only to protocols of quantum cryptography, but also to quantum computation, which is our future task.

\section*{Data availability statement}
The data obtained by numerical calculations and C++ programs will be available from the author upon reasonable request.

\section*{Conflict of interest}
The author declares no conflict of interest.

\section*{ORCID iDs}
Hiroo Azuma https://orcid.org/0000-0002-4374-8727
\\

\appendix
\section{\label{section-derivation-equations}Details of derivations of some equations}
\subsection{\label{subsection_Bdagger-Sa-B-0}
The derivation of equation~(\ref{Bdagger-Sa-B-0})}
From equations~(\ref{transformation-B-delta-0}) and (\ref{Sa-Da-definition-0})
and
\begin{equation}
\hat{B}^{\dagger}(\delta)\hat{a}^{\dagger}\hat{B}(\delta)
=
\frac{1}{\sqrt{2}}(\hat{a}^{\dagger}+e^{-i\delta}\hat{b}^{\dagger}),
\label{Bdagger-a-B-formula-0}
\end{equation}
we obtain
\begin{eqnarray}
\hat{B}^{\dagger}(\delta)\hat{S}_{a}(-\zeta)\hat{B}(\delta)
&=&
\exp[\frac{\zeta}{4}(\hat{a}^{\dagger}+e^{-i\delta}\hat{b}^{\dagger})^{2}
-
\frac{\zeta^{*}}{4}(\hat{a}+e^{i\delta}\hat{b})^{2}] \nonumber \\
&=&
\exp[\frac{1}{4}(\zeta\hat{a}^{\dagger 2}-\zeta^{*}\hat{a}^{2})
+
\frac{1}{4}(\zeta e^{-2i\delta}\hat{b}^{\dagger 2}-\zeta^{*}e^{2i\delta}\hat{b}^{2}) \nonumber \\
&&
+
\frac{1}{2}(\zeta e^{-i\delta}\hat{a}^{\dagger}\hat{b}^{\dagger}
-
\zeta^{*}e^{i\delta}\hat{a}\hat{b})].
\label{B-Sa-B-formula-0}
\end{eqnarray}
Because of
\begin{eqnarray}
&&
[\zeta\hat{a}^{\dagger 2}-\zeta^{*}\hat{a}^{2},
\zeta e^{-i\delta}\hat{a}^{\dagger}\hat{b}^{\dagger}-\zeta^{*} e^{i\delta}\hat{a}\hat{b}] \nonumber \\
&&
+
[\zeta e^{-2i\delta}\hat{b}^{\dagger 2}-\zeta^{*} e^{2i\delta}\hat{b}^{2},
\zeta e^{-i\delta}\hat{a}^{\dagger}\hat{b}^{\dagger}-\zeta^{*} e^{i\delta}\hat{a}\hat{b}] \nonumber \\
&=&
0,
\end{eqnarray}
we reach equation~(\ref{Bdagger-Sa-B-0}).

\subsection{\label{subsection_Bdagger-Da-B-0}
The derivation of equation~(\ref{Bdagger-Da-B-0})}
From equations~(\ref{transformation-B-delta-0}) and (\ref{Sa-Da-definition-0}), and (\ref{Bdagger-a-B-formula-0}),
we obtain
\begin{eqnarray}
\hat{B}^{\dagger}(\delta)\hat{D}_{a}(\alpha)\hat{B}(\delta)
&=&
\hat{B}^{\dagger}(\delta)
\exp(\alpha\hat{a}^{\dagger}-\alpha^{*}\hat{a})
\hat{B}(\delta) \nonumber \\
&=&
\exp[\frac{\alpha}{\sqrt{2}}(\hat{a}^{\dagger}+e^{-i\delta}\hat{b}^{\dagger})
-
\frac{\alpha^{*}}{\sqrt{2}}(\hat{a}+e^{i\delta}\hat{b})],
\end{eqnarray}
thus we can attain equation~(\ref{Bdagger-Da-B-0}).

\subsection{\label{subsection_Cna-Snma-Dma}
The derivations of equations~(\ref{Cna-0}), (\ref{Snma-0}), and (\ref{Dma-0})}
From equation~(\ref{Sa-Da-0state}),
we obtain
\begin{eqnarray}
C_{n}^{a}
&=&
{}_{a}\langle n|\hat{S}_{a}(-r/2)\hat{D}_{a}(\alpha/\sqrt{2})|0\rangle_{a} \nonumber \\
&=&
\sum_{m=0}^{\infty}
{}_{a}\langle n|\hat{S}_{a}(-r/2)|m\rangle_{a}
{}_{a}\langle m|\hat{D}_{a}(\alpha/\sqrt{2})|0\rangle_{a} \nonumber \\
&=&
\sum_{n=0}^{\infty}
S_{nm}^{a}D_{m}^{a},
\end{eqnarray}
where
\begin{eqnarray}
S_{nm}^{a}
&=&
{}_{a}\langle n|\hat{S}_{a}(-r/2)|m\rangle_{a}, \nonumber \\
D_{m}^{a}
&=&
{}_{a}\langle m|\hat{D}_{a}(\alpha/\sqrt{2})|0\rangle_{a}.
\end{eqnarray}
Thus, we reach equation~(\ref{Cna-0}).

First, we derive an explicit form of $S_{nm}^{a}$.
Using the following formulae:
\begin{eqnarray}
\hat{S}_{a}(\zeta)
&=&
\exp[-\frac{1}{2}\exp(i\varphi)\tanh r\hat{a}^{\dagger 2}] \nonumber \\
&&
\times
\exp[-\frac{1}{2}\ln(\cosh r)(\hat{a}^{\dagger}\hat{a}+\hat{a}\hat{a}^{\dagger})] \nonumber \\
&&
\times
\exp[\frac{1}{2}\exp(-i\varphi)\tanh r\hat{a}^{2}],
\end{eqnarray}
where $\zeta=r\exp(i\varphi)$,
\begin{eqnarray}
&&
\exp[-\frac{1}{2}\tanh(r/2)\hat{a}^{2}]|m\rangle_{a} \nonumber \\
&=&
\sum_{k=0}^{\lfloor m/2\rfloor}
\frac{1}{k!}
[-\frac{1}{2}\tanh(r/2)]^{k}
\sqrt{\frac{m!}{(m-2k)!}}|m-2k\rangle,
\end{eqnarray}
\begin{eqnarray}
&&
{}_{a}\langle n|\exp[\frac{1}{2}\tanh(r/2)\hat{a}^{\dagger 2}] \nonumber \\
&=&
\sum_{l=0}^{\lfloor n/2\rfloor}
\frac{1}{l!}
[\frac{1}{2}\tanh(r/2)]^{l}
\sqrt{\frac{n!}{(n-2l)!}}\langle n-2l|,
\end{eqnarray}
\begin{equation}
(\hat{a}^{\dagger}\hat{a}+\hat{a}\hat{a}^{\dagger})|n\rangle
=
(1+2n)|n\rangle,
\end{equation}
we obtain equation~(\ref{Snma-0}).

Second, we derive an explicit form of $D_{m}^{a}$.
Using the formula,
\begin{equation}
\hat{D}(\alpha)
=
\exp(-\frac{|\alpha|^{2}}{2})
\exp(\alpha\hat{a}^{\dagger})
\exp(-\alpha^{*}\hat{a}),
\end{equation}
we obtain
\begin{eqnarray}
D_{m}^{a}
&=&
{}_{a}\langle m|\hat{D}_{a}(\alpha/\sqrt{2})|0\rangle_{a} \nonumber \\
&=&
\exp(-|\alpha|^{2}/4)
{}_{a}\langle m|\sum_{k=0}^{\infty}\frac{1}{k!}(\frac{\alpha}{\sqrt{2}})^{k}\sqrt{k!}|k\rangle_{a} \nonumber \\
&=&
\exp(-|\alpha|^{2}/4)\frac{1}{\sqrt{m!}}(\frac{\alpha}{\sqrt{2}})^{m},
\end{eqnarray}
thus we attain equation~(\ref{Dma-0}).

\subsection{\label{subsection_Sn1n2lkab}
The derivation of equation~(\ref{Sn1n2lkab-01})}
Using the following formulae:
\begin{eqnarray}
\hat{S}_{ab}(\zeta)
&=&
\exp[-\exp(i\varphi)\tanh r\hat{a}^{\dagger}\hat{b}^{\dagger}] \nonumber \\
&&
\times
\exp[-\ln(\cosh r)(\hat{a}^{\dagger}\hat{a}+\hat{b}\hat{b}^{\dagger})] \nonumber \\
&&
\times
\exp[\exp(-i\varphi)\tanh r\hat{b}\hat{a}],
\end{eqnarray}
where $\zeta=r\exp(i\varphi)$,
\begin{eqnarray}
&&
\exp[-\tanh (r/2)\hat{b}\hat{a}]|m,k\rangle \nonumber \\
&=&
\sum_{l=0}^{\mbox{\scriptsize min}(m,k)}
(-1)^{l}\tanh (r/2)\frac{1}{l!}
\sqrt{\frac{m!k!}{(m-l)!(k-l)!}}|m-l,k-l\rangle,
\end{eqnarray}
\begin{equation}
(\hat{a}^{\dagger}\hat{a}+\hat{b}\hat{b}^{\dagger})|n,m\rangle
=
(n+m+1)|n,m\rangle,
\end{equation}
\begin{eqnarray}
&&
\langle n_{1},n_{2}|
\exp[\tanh (r/2)\hat{a}^{\dagger}\hat{b}^{\dagger}] \nonumber \\
&=&
\sum_{j=0}^{\mbox{\scriptsize min}(n_{1},n_{2})}
\tanh^{j}(r/2)\frac{1}{j!}
\sqrt{\frac{n_{1}!n_{2}!}{(n_{1}-j)!(n_{2}-j)!}}
\langle n_{1}-j,n_{2}-j|,
\end{eqnarray}
we obtain equation~(\ref{Sn1n2lkab-01}).
\\

\end{document}